\documentstyle[aps,prd,eqsecnum,epsf]{revtex}
\draft
\begin{document}
\wideabs{
%
\title
{Possible direct method to determine the radius of a star \break
from the spectrum of gravitational wave signals II : 
Spectra for various cases}
%
\author
{Motoyuki Saijo
\thanks{Electronic address :
saijo@astro.physics.uiuc.edu}
}
%
\address
{Department of Physics, University of Illinois at Urbana-Champaign,
1110 West Green Street, Urbana, IL 61801-3080}
%
\author
{Takashi Nakamura
\thanks{Electronic address : takashi@yukawa.kyoto-u.ac.jp}
}
%
\address
{Yukawa Institute for Theoretical Physics, Kyoto University,
Kyoto 606-8502, Japan}
%
\date{Received 27 August 2000}
\maketitle
%
\begin{abstract}
We compute the spectrum and the waveform of gravitational waves 
generated by the inspiral of a disk or a spherical like dust body into 
a Kerr black hole.  We investigate the effect of the radius R of the 
body on gravitational waves and conclude that the radius is 
inferred from the gravitational wave signal irrespective of (1) the 
form of the body (a disk or a spherical star) (2) the location where 
the shape of the body is determined, (3) the orbital angular momentum
of the body, and (4) a  black hole rotation.  We find that when $R$ is
much larger than the  characteristic length of the quasinormal mode
frequency, the spectrum  has several peaks and the separation of the
troughs $\Delta\omega$ is proportional to $R^{-1}$.  Thus, we may
directly determine the radius of a star in a coalescing binary black
hole - star system from the observed spectrum of gravitational
waves.  For example, both trough frequency of neutron
stars and white dwarfs are within the detectable frequency range of
some laser interferometers and resonant type detectors so that this
effect can be observed in the future.  We therefore conclude that the
spectrum of gravitational waves  may provide us important
signals in gravitational wave astronomy as in optical
astronomy.
\end{abstract}
}
%
%
\section{Introduction}
%

Researchers in gravitational physics have been expecting for the direct
detection of gravitational waves (GW) using laser interferometers
such as LIGO, VIRGO, GEO600, TAMA300, and  LISA as well as several
projects using resonant type detectors \cite{Thorne}.  The direct 
detection of GW will provide us not only the verification of General
Relativity  but also plenty of new aspects in many other fields, such
as  nonlinear physics related to  the Einstein  equations, nuclear and
particle physics from the estimation of equation of states of the
neutron star (NS), cosmology from the evidence of inflationary
universe and from the measurements of cosmological parameters. 
Therefore, it will open the frontiers of physics, gravitational wave
physics/astronomy in the 21st century.

Among  many possible sources of GW, coalescing binaries composed
of black holes (BHs) or NSs are probably the most  promising.  From
these gravitational wave signals, we can trace the astrophysical event.
Comparing the inspiral waveform with the theoretical templates using
the matched filtering technique, we may determine masses and spins of
the BH and star, respectively \cite{Thorne}.  On the other hand,
different type of information from the inspiral phase may be obtained
from the final merging phase such as the radius of a star, the equation
of state of a high density matter, and the structure of the stars.  Since
many physical elements have to be taken into account at once to obtain
such information in the merging phase, little numerical and  analytical 
study have been effectively investigated.  Recently, Vallisneri
\cite{Vallisneri} argued a possibility to extract the radius of a star
from GW.  If we measure the frequency of GW from the location where
the star is tidally disrupted, we may find information about the central
density of the star, and from that, we may obtain the radius of a
star if we know its equation of state.  This method as well as  other
methods using the results of numerical simulations, however, may
need information about the equation of state to determine the
radius of the star. For example, Vallisneri \cite{Vallisneri} used the
result of an incompressible homogeneous Newtonian fluid in a Kerr BH
\cite{Shibata}.

The first step to tackle this problem of the merging phase  might be
examined by using a BH perturbation approach.  In this paper, we shall
consider GW from either an usual star or a NS inspiraling into a BH,
emphasizing on the size effect of a disk/star on GW using a BH
perturbation approach. Many papers along this approach have been
published on this topic; GW from a  test particle plunging into a BH,
since Sasaki and Nakamura \cite{SN82a,SN82b} reformulated Teukolsky
equation \cite{Teukolsky} to make it possible to calculate GW induced
by a particle plunging into a Kerr BH from infinity.  Especially, several
papers focus on phase cancellation effects on GW using this
approach. In early studies by Nakamura and Sasaki
\cite{NS} and Haugan, Shapiro, and Wasserman \cite{HSW}, GW from a
deformed shell falling into a Schwarzschild BH were investigated. 
Saijo, Shinkai, and Maeda \cite{SSM} extended their model to a Kerr BH
and found  that when the central BH is rotating,  a slightly oblate dust
shell minimizes the collapsing energy of GW but with a non-zero finite
value, which depends on the Kerr parameter.

A BH perturbation approach is not only a toy model to illustrate GW
from a binary system.  There are two well-known facts comparing the
results from the BH perturbation approach with the full numerical
simulation.  The research of the head on collision of two BHs which
was  pioneered by Smarr \cite{Smarr} and updated by Anninos, Price,
Pullin, Seidel, and Suen \cite{APPSS} was the first work.  They
compared their numerical results with that of the BH perturbation
approach ; GW from a test particle falling into a Schwarzschild BH by
Davis, Ruffini, Press, and Price \cite{Davis}.  The radiated energy
agrees with each other within a factor of two.  Especially, the
waveform of the ringing tail has a very accurate coincidence between
these two cases.  The second research was an axisymmetric rotating
star collapse by Stark and Piran \cite{Stark}. They also compared their
result with the previous semi-numerical result; ring shaped test
particles plunging into a Kerr BH by  Nakamura, Oohara, and Kojima
\cite{NOK}.  Although these two models are quite different, the
waveforms of these two computations give a fairly good agreement. 
We can interpret the reason why these different two
results have a coincidence as follows.  The main body of the rotating
star first collapses to form a Kerr BH.  Then, the remaining part which
is essentially regarded as ring-shaped particles fall into the newly
formed BH.  Therefore, the above agreement is not so strange from this
physical picture of a rotating star collapse.

These two examples encourage us to apply the BH perturbation approach
to the final phase of coalescing binary systems.   The model worked in
this approach may throw a milestone on what kind of situation provides
us a very impressive viewpoint for GW and needs to be followed by the
detailed, numerical simulations.  The result from this approach also
helps those who work in this field to concentrate on such predicted,
exciting results.

Our purpose in this paper is to focus on the size effect of a star on
GW from a coalescing binary BH - star system.  We use the BH
perturbation approach for the sake of the detailed, perspective
investigation to find the property appears in the GW.  In Ref.
\cite{SN00}, we have already presented the result with a limited case;
a disk with one set of parameters.  In this paper, we examine the
dependence of the results on (1) the form of the body (a disk or a
spherical star), (2) the location where  the shape of the body is
determined, (3) the angular momentum of the body, and (4) a BH
rotation in detail.

This paper is organized as follows. In Sec. \ref{sec:WaveFunction},  we
introduce the method how to construct the radial wave function of GW
for  disk shaped test particles as well as spherical shaped ones using
the BH perturbation approach.  We present our numerical results of GW
from a disk and a spherical dust star in Sec. \ref{sec:GW}.  Section
\ref{sec:Discussion} is devoted to discussion.  Throughout this paper,
we use the geometrized units of $G=c=1$ and the metric sign as
$(-,+,+,+)$.

%
\section{Radial wave function }
\label{sec:WaveFunction}
%

In this section, we explain the method to construct the radial wave
function of a dust disk and a spherical star in Kerr spacetime using the
BH perturbation approach.  Before explaining the construction method
of the radial wave function, let us briefly introduce the BH
perturbation approach
\cite{SN82a,SN82b,NOK}. This approach is applicable only  if $M \gg
\mu$, where $M$ is the BH mass and $\mu$ is the particle mass.
However, the computational result happens to be well-behaved in some
cases even if we extrapolate the result to $\mu \sim M$ as we have
already mentioned before.  In order to extract GW at infinity with
sufficient computational accuracy,  we have to solve the generalized
Regge-Wheeler equation  (Sasaki-Nakamura equation) [Eq. (2.28) of Ref.
\cite{SN82b}] such as
\begin{equation}
\left[ \frac{d^{2}}{dr^{*2}} - F(r^{*})\frac{d}{dr^*} - U(r^{*})  \right]
X_{lm\omega}(r^{*}) = S_{lm\omega}(r^{*}),
\label{eqn:GeneRW}
\end{equation}
where $r^{*}$, $S_{lm\omega}(r^{*})$, $F(r^{*})$ and $U(r^{*})$ are the
tortoise coordinate of a Kerr BH, the source term from a test particle
of mass $\mu$ [Eq. (2.29) of Ref. \cite{SN82b}] and two potential
functions [Eqs. (2.12a) and (2.12b) of Ref. \cite{SN82b}], respectively.
To solve $X_{lm\omega}(r^{*})$  using a Green function method, we need
two independent homogeneous solutions  whose boundary conditions
are given by
\begin{eqnarray*}
X_{lm\omega}^{\rm in(0)}(r^{*}) & = & \left\{
\begin{array}{ll}
e^{-i k r^{*}} &  r^{*}\rightarrow -\infty \\
A_{lm\omega}^{\rm in}e^{-i \omega r^{*}} + 
A_{lm\omega}^{\rm out}e^{i \omega r^{*}} &  
r^{*}\rightarrow \infty
\end{array}
\right. ,\\
X_{lm\omega}^{\rm out(0)}(r^*) & = & \left\{
\begin{array}{ll}
B_{lm\omega}^{\rm in} e^{-i k r^{*}} + 
B_{lm\omega}^{\rm out}e^{i k r^{*}} &
r^{*}\rightarrow - \infty \\
e^{i \omega r^{*}} & r^{*}\rightarrow \infty
\end{array}
\right. ,
\end{eqnarray*}
where $k=\omega-ma/[2(M+\sqrt{M^2-a^2})]$.
The inhomogeneous solution to Eq. (\ref{eqn:GeneRW}) becomes
\begin{eqnarray*}
X_{lm\omega}(r^*) & = &
X_{lm\omega}^{\rm in(0)}
\int_{r^{*}}^{\infty} \frac{S_{lm\omega}(r^*)}{W}
X_{lm\omega}^{\rm out(0)}~dr^{*}
\nonumber \\
&&
+ X_{lm\omega}^{\rm out(0)}
\int_{-\infty}^{r^{*}} {S_{lm\omega}(r^*)\over {W}}
X_{lm\omega}^{\rm in(0)}~dr^{*},
\end{eqnarray*}
where $W$ is the Wronskian,
\begin{equation}
W  \equiv X_{lm\omega}^{\rm in(0)}
\frac{dX_{lm\omega}^{\rm out(0)}}{dr^{*}} -
X_{lm\omega}^{\rm out(0)}
\frac{dX_{lm\omega}^{\rm in(0)}}{dr^{*}}
.
\end{equation}
The asymptotic behavior  of the radial wave function
$X_{lm\omega}(r^{*})$ is given by
\begin{eqnarray}
X_{lm\omega}(r^{*}) & = &
A_{lm\omega} e^{i\omega r^{*}},
\label{eqn:RWsol}\\
A_{lm\omega} & = & \int_{-\infty}^{\infty}
\frac{S_{lm\omega}(r^{*})}{W} X_{lm\omega}^{\rm in(0)}~dr^{*}
. \nonumber
\end{eqnarray}
Using the amplitude of the radial wave function, the energy spectrum
and the waveform of GW [Eqs. (3.6) and (3.10) of Ref. \cite{SSM}] are
given by
\begin{eqnarray}
\left( \frac{dE}{d\omega} \right)_{lm\omega} & = &
8 \omega^{2}
\left| \frac{A_{lm\omega}}{c_{0}} \right|^{2}
,
\hspace{1cm}
(- \infty < \omega < \infty)
\label{eqn:dEdw}
,\\
h_{+}-ih_{\times} &=&
\frac{8}{r} \int_{-\infty}^{\infty} d\omega
e^{i \omega (r^{*} - t)}
\nonumber \\
&& \times
\sum_{l,m} \frac{A_{lm\omega}}{c_{0}}
\mbox{}_{-2}S_{lm}^{a\omega}(\theta)
\frac{e^{i m \varphi}}{\sqrt{2\pi}},
\label{eqn:Waveform}
\end{eqnarray}
where $r$ is the coordinate radius from the center of the BH,
$\mbox{}_{-2}S_{lm}^{a\omega}$ is spin $- 2$ weighted spherical
harmonics, $c_{0}$ is a constant given in Ref. \cite{SSM}.

First, we consider a dust disk plunging into a
Kerr BH from infinity.  We set three assumptions to construct a disk
in Kerr spacetime.  (1) The disk is made up of test particles and
each of them moves in the equatorial plane plunging into the Kerr BH
from infinity.  (2) The shape of the disk is set at the location 
$r=r_{0}$.  (3) All component particles of the disk have same
energy, angular momentum, and Carter constant.  In this paper, we
set the disk at the location $r_{0}$ as $10M$ or $20M$, because our
purpose in this paper is to focus on the merging phase of the coalescing
binary BH - star \cite{ODE}.  The motion of a test particle in Kerr
spacetime is written in general as \cite{Carter68}
\begin{eqnarray}
\Sigma \frac{dt}{d\tau} &=&
- a ( a \tilde{E} \sin^{2}\theta - \tilde{L}_{z} ) +
\frac{r^{2} + a^{2}}{\Delta} P
,\\
\Sigma \frac{dr}{d\tau} &=& \pm \sqrt{R}
,\\
\Sigma \frac{d\theta}{d\tau} &=& \pm \sqrt{\Theta}
,\\
\Sigma \frac{d\phi}{d\tau} &=&
- \left( a \tilde{E} - \frac{\tilde{L}_{z}}{\sin^{2}\theta} \right) +
\frac{a}{\Delta} P
,
\end{eqnarray}
where $E = \mu \tilde{E}$ is the energy, $L_{z} = \mu \tilde{L}_{z}$ is
the orbital angular momentum, $C$ is Cater constant of the particle,
$\Sigma = r^{2} + a^{2} \cos^{2}\theta$, $\Delta = r^{2} - 2 M r + a^{2}$. 
The symbols $P$, $R$ and $\Theta$ are defined by
\begin{eqnarray}
P &=&
\tilde{E} (r^{2} + a^{2} ) - a \tilde{L}_{z}
,\\
R &=&
P^{2} - \Delta [ r^{2} + ( \tilde{L}_{z} - a \tilde{E} )^{2} + C ]
,\\
\Theta &=&
C - \cos^{2}\theta
\left[ a^{2} ( 1 - \tilde{E}^{2} ) +
\frac{\tilde{L}_{z}^{2}}{\sin^{2}\theta} \right]
.
\end{eqnarray}
When a particle moves in the equatorial plane or in the constant
zenith angle, Carter constant is required to be zero from the
stability of its trajectory.

Here we explain our effective method to construct the radial wave
function, instead of computing all the trajectories of each component
particle.  We note that the geodesic in the equatorial plane of a  Kerr
BH is characterized by two parameters; the specific energy
($\tilde{E}$) and the specific angular momentum ($\tilde{L}_{z}$).  Let
$t=T(r)$ and $\phi=\Phi(r)$ express the orbit of the geodesic for given
$\tilde{E}$ and $\tilde{L}_{z}$.  Since there is  time symmetry and
azimuthal symmetry in Kerr spacetime, we can find another geodesic
with same energy and angular momentum.  Then, we can set another
geodesic for same $\tilde{E}$ and $\tilde{L}_{z}$ with shifting time
and azimuthal angle as $t  = T(r) + c_{t}$ and $\phi = \Phi(r) +
c_{\phi}$ where $c_{t}$ and $c_{\phi}$ are constants.  After we set the
point where the representative particle pass through ($r = r_{i}$ and
$\phi = \phi_{i}$ at $t = T(r_{0})$),  the orbit of each component
particle can be expressed  as $t = T(r) + T(r_{0}) - T(r_{i})$ and $\phi =
\Phi(r) + \phi_{i} - \Phi(r_{i})$. Therefore, it becomes possible to set a
number of particles with same $\tilde{E}$ and $\tilde{L}_{z}$ to form
the disk of radius $R$ whose center is $r=r_{0}$ and $\phi = \Phi
(r_{0})$ at $t = T(r_{0})$.  Using this character, we can construct the
radial wave function only with a function of $r$ 
\cite{NS,HSW,SSM}.  We can also choose arbitrary  $\tilde{E}$ and
$\tilde{L}_{z}$ for the disk except for the condition that each test
particle plunges into a Kerr BH.  The parameter range of 
$\tilde{L}_{z}$ in the case of $\tilde{E} = 1$ which we use in this
paper is
\begin{equation}
-2M - \sqrt{4M^{2} + 4 M a}
< \tilde{L}_{z}  <
2M + \sqrt{4M^{2} - 4 M a}.
\label{eqn:Lzcondition}
\end{equation}

We construct the radial wave function of GW for a dust disk.  Since
Kerr spacetime has a property of symmetry for $t$ and $\phi$, we can
use this symmetry to construct the radial wave function in order to
save computational time as we explained above.  Therefore, the radial
wave function of the  disk $A_{lm\omega}^{\rm (disk)}$ can be
constructed from that of a test particle $A_{lm\omega}^{\rm
(particle)}$ as
\begin{eqnarray}
A_{lm\omega}^{\rm (disk)}
&=& f_{m\omega} A_{lm\omega}^{\rm (particle)}
,
\label{eqn:DiskRadialFunctionSeparation}
\\
f_{m\omega}
&=&
2 \frac{\mu}{S}
\int_{r_{0} - R}^{r_{0} + R} dr r  \frac{\sin(m \phi_{0} (r))}{m}
\nonumber \\
&&
\times
e^{i [ \omega [ t(r) - t(r_{0}) ] -
m [ \phi(r) - \phi(r_{0}) ] ]},
\label{eqn:FormFactor}
\\
\phi_{0}(r) &=&
\cos^{-1} \frac{r^{2}+r_{0}^{2}- R^{2}}{2rr_{0}},
\end{eqnarray}
where $\mu$ is the mass of the disk, $r_{0}$ is the location of the
central point of the disk where we set the circular disk, $R$ is the
coordinate radius of the disk at
$r=r_{0}$,  $S$ is the normalization factor defined by
\begin{eqnarray}
S &=&  \int_{S} ds =
2 \int_{r_{0} - R}^{r_{0} + R} dr r \phi_{0} (r),
\label{eqn:sphere}
\\
\phi_{0}(r) &=&
\cos^{-1} \frac{r^{2}+r_{0}^{2}- R^{2}}{2rr_{0}}
.
\end{eqnarray}
In order to construct $A_{lm\omega}^{\rm (disk)}$ [Eq.
(\ref{eqn:DiskRadialFunctionSeparation})], we only have to prepare
$A_{lm\omega}^{\rm (particle)}$,  $t(r)$, and $\phi(r)$.

Next, we present how to set a spherical dust star at $r=r_{0}$
($r_{0}=10M$, $20M$) in order to set a realistic model of BH - star
merging.  We also set three assumptions to construct a spherical
dust star in Kerr spacetime. (1) The star is made up of test particles
that move along the constant zenith angle of a Kerr BH. (2) The shape of
the star is set at the location of $r=r_{0}$, and the center of the star
is located in the equatorial plane of a Kerr BH. (3) All test
particles have the same specific energy, specific orbital angular
momentum, and Carter constant.  From the assumptions (1) and (3), we
cannot take the orbital angular momentum of the particles into
consideration.  Since the orbital angular momentum has almost lost by
the gravitational wave emission in the previous inspiral phase, the
zero orbital angular momentum may not be an absurd approximation. 
Our main purpose in this modeling to construct a spherical star is to
confirm that the difference of the body (a disk or a spherical star)
little affect our main conclusion.  We expect this proposition is
true because the size effect in our model does not depend on the orbit
of the star. We also note that the stability of the constant
zenith angle  trajectory also requires
$C=0$ when we choose $\tilde{E} = 1$.

We use the same technique as the dust disk case to construct the
radial wave function of the spherical dust star.  We can only adopt this
technique to the particles in each constant zenith angle, the radial
wave function for the spherical dust star $A_{lm\omega}^{\rm (star)}$
becomes a function of $\theta$ as
\begin{eqnarray}
A_{lm\omega}^{\rm (star)}
&=&
2 \frac{\mu}{V}
\int_{r_{0} - R}^{r_{0} + R} dr r^{2}  \frac{\sin (m\phi (r,\theta))}{m}
\nonumber \\
&&
\times
\int_{-\theta(r)}^{\theta(r)} d\theta \sin\theta
A_{lm\omega}^{\rm (particle)} (\theta)
\nonumber \\
&&
\times
e^{i [ \omega [ t(r, \theta) - t(r_{0}, \theta) ] -
m [ \phi(r, \theta) - \phi(r_{0}, \theta) ] ]},
\label{eqn:StarRadialFunction}
\end{eqnarray}
where $\mu$ is a mass of the star, $r_{0}$ is the location of the
central point of the spherical dust star where we set $r_{0}=10M$,
$20M$ in this paper, $R$ is a coordinate radius of the star, $V$ is 
the normalization constant of the star defined by
\begin{eqnarray}
V &=&  \int_{V} dv =
2 \int_{r_{0} - R}^{r_{0} + R} dr r^{2} \phi (r,\theta)
\nonumber \\
&& \times
\int_{-\theta(r)}^{\theta(r)} d\theta \sin\theta ,
\label{eqn:volume}
\\
\phi (r,\theta) &=&
\cos^{-1} \frac{r^{2}+r_{0}^{2} - R^{2}}
{2rr_{0}\sin\theta (r)}
,\\
\theta(r) &=&
\cos^{-1} \frac{r^{2}+r_{0}^{2}- R^{2}}{2rr_{0}}
.
\end{eqnarray}
Therefore, we have to prepare $A_{lm\omega}^{\rm
(particle)}(\theta)$ and $t(r, \theta)$, $\phi(r, \theta)$
for a set of constant zenith angle trajectories in order to construct
$A_{lm\omega}^{\rm (star)}$  [Eq. (\ref{eqn:StarRadialFunction})].

%
\section{Gravitational waves from the disk and the spherical star}
\label{sec:GW}
%
%
We show our numerical results for a dust disk plunging into a
Kerr BH.  We calculate GW for a wide range of parameters
($\tilde{L}_{z}/M=2$, $0$, $-3$, $a/M=0$, $0.5$, $0.9$, $r_{0}=10M$,
$20M$) than before \cite{SN00}.  We place the spectra into three
categories, classified by the size of the disk.

We briefly explain the feature of each type of the spectrum.
Type 1 spectrum has the same character as that of a test particle;  the
energy spectrum has one dominant characteristic frequency for each
$m$ mode where most of GW are radiated.  This frequency corresponds
to the quasinormal mode (QNM) of the background BH.  The
real part corresponds to the resonant oscillation of the BH, and the
imaginary part corresponds to the ``damping" time scale of GW at the
late time. The waveform, which has a damping oscillation at the late
time, can  also be described by the QNM frequency at that part.  
Since GW from a disk has a very similar character to that from a test
particle, we may ignore the size effect on GW as far as the diameter
of the disk is smaller than the characteristic length of the QNM
frequency.

We classify the spectrum to type 2 when the diameter of the disk is
comparable to the  characteristic length of the  QNM frequency.  In this
case, the sharp peak near the QNM frequency has been weakened.   GW
still have a damping oscillation in the waveform, though the amplitude
of the ringing tail is much smaller than that for a test particle due to 
the phase cancellation effect on GW.

When the diameter of the disk is larger than the characteristic length
of the QNM frequency,  the spectrum has several peaks (type 3).  One
remarkable feature appears in the spectrum that the separation of the
troughs does not largely depend on the orbit.  Using this character, we
find a direct method to determine the radius of a star \cite{SN00}.  In
the waveform, we can hardly recognize the damping oscillation at the
late time.  Therefore, we cannot find any characteristic frequency of
GW, corresponds to the QNM frequency for a test particle in a plunging
orbit.

\subsection{Dependence on a BH rotation}
\label{subsec:BHrotation}
First, we discuss the dependence of the energy spectra and
the waveforms on the radius of the disk and the BH rotation, keeping
the other set of parameters as $r_{0}=10M$, $\tilde{L}_{z} / M = 2$. 
Since we choose $\tilde{L}_{z}/M = 2$, we only focus on $l=m=2$
nonaxisymmetric mode which is the dominant mode in all modes in
both spectra and waveforms.

For a test particle case, the peak frequency which corresponds to
the QNM frequency of the BH is different among these three spectra
[Figs. \ref{fig:r10a0z2sp} (a), \ref{fig:r10a5z2sp} (a), and
\ref{fig:r10a9z2sp} (a)] .  For example, the spectrum has a peak at
frequency $M\omega = 0.33$ ($l=m=2$) in Fig. \ref{fig:r10a0z2sp} (a)
($a=0$),  frequency $M\omega = 0.41$ ($l=m=2$) in Fig.
\ref{fig:r10a5z2sp} (a) ($a/M=0.5$),  frequency $M\omega = 0.63$
($l=m=2$) in Fig. \ref{fig:r10a9z2sp} (a) ($a/M=0.9$), while QNM
frequency for $a=0$, is $M \omega = 0.37 - 0.089 i$ ($l=2$), for
$a/M=0.5$ is $M \omega = 0.46 - 0.085 i$ ($l=m=2$), and for $a/M=0.9$
is $M \omega = 0.66 - 0.065 i$ ($l=m=2$) [Fig. 1a of Ref.
\cite{Detweiler} and Fig. 3 (c) of Ref. \cite{Leaver}].   These results
indicate the fact that the vibration of the BH has a dominant effect on
GW for a test particle in a  plunging orbit.  When we turn to look at
waveforms, Figs. \ref{fig:r10a0z2gw} (a), \ref{fig:r10a5z2gw} (a), and
\ref{fig:r10a9z2gw} (a) have  damping oscillations at the late time,
which can be described by the QNM frequency (Fig. 4 of Ref. \cite{KN1}
and Fig. 4 of Ref. \cite{KN2}).

For type 1 case, both energy spectra and waveforms for a disk have a
very similar behavior to the case for a test particle.  The small size of
the disk (for example $R=0.785M$) does not largely affect energy
spectra and waveforms.  The characteristic behavior in spectra and 
waveforms is almost the same as the case of a test particle.  We 
confirm this feature by comparing  Fig. \ref{fig:r10a0z2sp} (Figs.
\ref{fig:r10a5z2sp}, \ref{fig:r10a9z2sp}) (a) and (b) for the spectra,
and Fig. \ref{fig:r10a0z2gw} (Figs. \ref{fig:r10a5z2gw},
\ref{fig:r10a9z2gw}) (a) and (b) for the waveforms.  The difference
between the spectrum from a test particle and that from a small size
disk appears only in the form factor as we follow the construction
of $A_{lm\omega}^{\rm (disk)}$ in Eq.
(\ref{eqn:DiskRadialFunctionSeparation}). When we look at the form
factor [Figs. \ref{fig:r10a0z2f} (a), \ref{fig:r10a5z2f} (a), and
\ref{fig:r10a9z2f} (a)], $|f_{2 \omega}|^{2}$ almost takes the value
unity for almost all range of the frequency, which is equivalent to the
case of a test particle.  The behavior of the form factor lead the
conclusion that there is little difference between $R=0.785 M$ and
$R=0$ from the viewpoint of gravitational wave signals in the plunging
orbit.

For type 2 case, both energy spectra and waveforms have a different
feature from those of a test particle.  In the energy spectrum, the
sharp peak near the QNM frequency has been weakened [Figs.
\ref{fig:r10a0z2sp} (c), \ref{fig:r10a5z2sp} (c), and
\ref{fig:r10a9z2sp} (c)].   Since we have defined the range of the
radius for type 2 that the diameter of the disk is comparable to the
characteristic length of the QNM frequency, the range
depends on the BH rotation.  We choose each range for each BH
rotation for type 2 case as $R/M\sim 2$ for  $a/M < 0.5$,
while $R/M\sim 1.5$ for $a/M=0.9$.  The difference of the radius for
each BH rotation is also confirmed from the form factor [Figs.
\ref{fig:r10a0z2f} (b), \ref{fig:r10a5z2f} (b), and \ref{fig:r10a9z2f}
(b)] since there is a sharp trough in the form factor appears near the
QNM frequency. When we turn to look at the waveform [Figs.
\ref{fig:r10a0z2gw} (c), \ref{fig:r10a5z2gw} (c), and
\ref{fig:r10a9z2gw} (c)], the amplitude near the merging phase has
been weakened due to the  phase cancellation effect on GW.  In spite of
this effect, we can still find a damping
oscillation quite clearly.

For type 3 case, both energy spectra and waveforms have a completely
different feature from the case of type 1 and type 2; several peaks
appear in the spectrum [Figs. \ref{fig:r10a0z2sp} (d),
\ref{fig:r10a5z2sp} (d), and \ref{fig:r10a9z2sp} (d)].  These peaks
also appear in the form factor [Figs. \ref{fig:r10a0z2f} (c),
\ref{fig:r10a5z2f} (c), and \ref{fig:r10a9z2f} (c)] and take the same
behavior to the spectrum.  We show the relation between the
separation frequency of the troughs and the radius of the disk in Table
\ref{tbl:EstimateR10}.  Remarkably, the relation
$R\Delta \omega \sim 1$ holds for a wide range of parameters.  This
relation make it possible to measure $R$ directly from the observed
$\Delta\omega$.  We will discuss  this point in more detail in Sec.
\ref{sec:Discussion}.  When we turn to look at the waveform [Figs.
\ref{fig:r10a0z2gw} (d), \ref{fig:r10a5z2gw} (d), and
\ref{fig:r10a9z2gw} (d)], we can hardly find a damping oscillation at
the late time due to the phase cancellation effect on GW.  We find
that the size effect only appears at the merging phase 
comparing the waveforms with each different radius.  For example, we 
find the difference within the range of $-100 < t-r^{*}/M < -30$ in Fig.
\ref{fig:r10a0z2gw} ($a=0$), within the range of $-100 < t-r^{*}/M <
-30$ in Fig. \ref{fig:r10a5z2gw} ($a/M=0.5$), within the range of
$-130 < t-r^{*}/M < -30$ in Fig. \ref{fig:r10a9z2gw} ($a/M=0.9$). 
Although we assume that all the component test particles come from
infinity, the waveforms in the merging phase might not depend on this
assumption so much.  Since the dominant wave is described by the
QNM frequency for a plunging orbit case, irrespective of the particle
energy and angular momentum,  the phase cancellation effect from
several QNM frequency waves is responsible to the behavior of the
waveforms in this merging phase.

\subsection{Dependence on an orbital angular momentum}
Next,  we discuss the dependence of energy spectra and
waveforms on the radius of the disk and the orbital angular momentum,
keeping the other set of  parameters as $r_{0}=10M$, $a/M = 0.9$.
For a test particle case,  the spectrum definitely depends on the
orbital angular momentum [Figs. \ref{fig:r10a9z2sp}(a),
\ref{fig:r10a9z263sp} (a), \ref{fig:r10a9z0sp} (a), and
\ref{fig:r10a9zm3sp} (a)].  In fact, GW are effectively radiated near
the QNM frequency and the spectrum takes an additional bump in
the case of $\tilde{L}_{z}/M = 2.63$ [Fig. \ref{fig:r10a9z263sp} (a)]. 
Since we do not take into account of the radiation reaction force
effect in our model, the amount of radiated energy diverges when the
particle takes a circular orbit.  Therefore, GW are radiated effectively
when the amount of the orbital angular momentum approaches to the
value of a circular orbit. In the case of $\tilde{L}_{z}/M=2$, $m=2$
mode dominates in $l=2$  spectrum while for $\tilde{L}_{z}/M=-3$,
$m=-2$ mode dominates in the spectrum within the positive frequency
range.  This tendency turns completely contrary within the negative
frequency range.  For example,
$m=-2$ mode dominates in the case of $\tilde{L}_{z}/M=2$, while
$m=2$ mode dominates for $\tilde{L}_{z}/M=-3$.  There is another
feature in the spectrum; each $\pm m$ mode has a reflection symmetry
at the zero frequency.  This comes from the fact that the
system has an equatorial plane symmetry.  When we turn to look at
waveforms (Figs.
\ref{fig:r10a9z2gw}, \ref{fig:r10a9z0gw}, and \ref{fig:r10a9zm3gw}),
some unclear damping oscillation appears in Figs. \ref{fig:r10a9z0gw}
(a) and \ref{fig:r10a9zm3gw} (a).  These behavior might come from the
superposition of several waves, each of them is described by a single
QNM frequency.

The classification of the type, depending on the radius of the disk, and
the behavior of each type are the same as the case of Subsec.
\ref{subsec:BHrotation}, we will briefly describe the results in each
category. Spectra, waveforms, and form factor for type 1 case are
shown in Figs.
\ref{fig:r10a9z2sp} (b), \ref{fig:r10a9z263sp} (b), \ref{fig:r10a9z0sp}
(b), \ref{fig:r10a9zm3sp} (b), in Figs. \ref{fig:r10a9z2gw} (b),
\ref{fig:r10a9z263gw} (b), \ref{fig:r10a9z0gw} (b),
\ref{fig:r10a9zm3gw} (b), and in Figs. \ref{fig:r10a9z2f} (b),
\ref{fig:r10a9z263f} (b), \ref{fig:r10a9z0f} (b), \ref{fig:r10a9zm3f}
(b).  Since this type little affects the size on GW, the disk can
almost be regarded as a test particle from the viewpoint of a
gravitational wave signal.  Spectra, waveforms, and form factor
for type 2 case are in Figs. \ref{fig:r10a9z2sp} (c),
\ref{fig:r10a9z263sp} (c), \ref{fig:r10a9z0sp} (c),
\ref{fig:r10a9zm3sp} (c), in Figs. \ref{fig:r10a9z2gw} (c),
\ref{fig:r10a9z263gw} (c), \ref{fig:r10a9z0gw} (c),
\ref{fig:r10a9zm3gw} (c), and in Figs. \ref{fig:r10a9z2f} (c),
\ref{fig:r10a9z263f} (c), \ref{fig:r10a9z0f} (c), \ref{fig:r10a9zm3f}
(c).  The characteristic QNM frequency, as appeared in the spectrum for
type 1 case, has been weakened due to the phase cancellation effect on
GW, but we can still look at the dumping oscillation in the waveform. 
Spectra, waveforms, and form factor for type 3 case are in Figs.
\ref{fig:r10a9z2sp} (d),
\ref{fig:r10a9z263sp} (d), \ref{fig:r10a9z0sp} (d),
\ref{fig:r10a9zm3sp} (d), in Figs. \ref{fig:r10a9z2gw} (d),
\ref{fig:r10a9z263gw} (d), \ref{fig:r10a9z0gw} (d),
\ref{fig:r10a9zm3gw} (d), and in Figs. \ref{fig:r10a9z2f} (d),
\ref{fig:r10a9z263f} (d), \ref{fig:r10a9z0f} (d), \ref{fig:r10a9zm3f}
(d).  We can hardly find the characteristic frequency in both spectra
and waveforms.  We also show the snapshot of the disk
form in Figs. \ref{fig:r10shapea9z2} and \ref{fig:r10shapea9z263},
focusing on the modification of the disk.  When we turn to look at $R
\Delta \omega \equiv C$ in Table \ref{tbl:EstimateR10}, the
dependence of the orbital angular momentum on $C$ is larger than that
of the BH rotation although it takes $0.87 < C < 1.16$ in any case.  The
value $C$ indeed depends on the orbit, as we will explain the structure
of $C$ in Sec. \ref{sec:Discussion} in detail.

\subsection{Dependence on the location where we set the circular disk}
Let us consider the effect of the radius and the location, where we
define the circular disk, on energy spectra and waveforms,
keeping the other set of parameters as $\tilde{L}_{z}/M=2$, $a/M =
0.9$.   The circular disk at $r_{0}=20M$ change the
shape and when it comes to $r=10M$ the aspect ratio
between the radial direction and azimuthal angle direction
becomes $3:1$ in the case of Fig. \ref{fig:r20shapea9z2}.  It is,
therefore, natural to investigate the dependence on $r_{0}$.  We
show the spectra, the waveform, and the form factor in the case of
$r_{0}=10M$ in Figs. \ref{fig:r10a9z2sp}, \ref{fig:r10a9z2gw}, and
\ref{fig:r10a9z2f} and in the case of $r_{0}=20M$ in Figs.
\ref{fig:r20a9z2sp}, \ref{fig:r20a9z2gw}, and \ref{fig:r20a9z2f}. 
Since  the spectra with the size range of $1.57 \lesssim R \lesssim
6.18$ have several peaks in $r_{0}=20M$ case, it may be
possible to determine the radius of the disk.  From Table
\ref{tbl:EstimateR20}, we have $R\Delta \omega\equiv C\sim 0.9$ 
for $r_{0}=20M$ case.  Since $C\sim 1$ for $r_{0}=10M$, the value of
$C$ slightly depends on $r_{0}$.  We can estimate the radius of the 
disk only using the separation of the peaks appeared in the spectrum. 
We will explain this point in Sec. \ref{sec:Discussion}.

For $r_{0}=20M$ case (Fig. \ref{fig:r20a9z2f}), there is one remarkable
difference in the form factor from $r_{0}=10M$ case (Fig.
\ref{fig:r10a9z2f}); the form factor almost has $\omega=0$
reflection symmetry for $l=m=2$ mode with the existence of the
orbital angular momentum.  This means that the system is almost
axisymmetry.  For $r_{0}=20M$ case, the late stage of the shape
turns out to be quite axisymmetry.  In fact, at the end of the evolution,
say $r \sim 2M$, the location of the particles approaches to
axisymmetric state (Fig. \ref{fig:r20shapea9z2}).  This causes less
gravitational wave emission and weakens the amplitude at the late
time.

\subsection{Spherical star}
\label{subsec:Star}
Finally, we show energy spectra and waveforms of GW from spherical
dust stars in the case of type 2 ($a/M=0.9$, $R=1.56M$) and type 3
($a/M=0.9$, $R=5.88M$).  We cannot express the radial wave functions
as the product of a test particle part and of a form factor part in the
spherical dust star, because $A_{lm\omega}^{\rm (particle)} (\theta)$,
$t (r,\theta)$, $\phi (r,\theta)$ depends on both $r$ and $\theta$. 
However, the main property of the spectrum from the spherical dust
star is the same as that from the disk.  When the radius of a
spherical dust star  is comparable to the characteristic length $L$ of
the QNM frequency, a phase cancellation effect on GW appears in the
spectrum [Fig. \ref{fig:starsp} (a)].  Therefore, the spectrum has a
different behavior from a test particle, which is the same to the disk
case as we mentioned in Subsec. \ref{subsec:BHrotation}.  When the 
radius of the spherical dust star is much larger than $L$,  several
peaks appear in the spectrum (Fig. \ref{fig:starsp} (b)).  The behavior
of the spectrum for a spherical star is similar to that for a disk case.  
We also show the dependence on $r_{0}$ in Fig. \ref{fig:starsp} (b),
(c).  These results have the same behavior to the disk case; the
spectrum has smaller amplitude for $r_{0}=20 M$ than for
$r_{0}=10M$. Since all of the characters appeared in the spectrum are very similar to the disk case, 
it is also possible for a spherical star to determine the radius
directly.  We summarize the estimation  of the radius in Table
\ref{tbl:EstimateRs10} and \ref{tbl:EstimateRs20} using the same
method as the disk case. The method still works quite well, but the
relative error rate becomes large when we compare the result
with disk case.  The reason for the large error may account for
$\theta$ dependence of the orbit.  Since all of the component particles
of the disk have the same $E$ and $L_{z}$ for the disk case, the
difference of the orbit comes only from $t$ and
$\phi$.  While, we have additional
$\theta$ dependence on the orbit for the dust star case, we cannot
estimate a simple relationship between $R$ and $\Delta \omega$ as
we will discuss in Sec. \ref{sec:Discussion}.  Therefore, the relative
error rate for the dust star case becomes larger than that for the dust
disk case.  In spite of the above reason, $R \Delta \omega$ is still
constant within 10\% error.  We also show the waveform of GW from
the spherical dust star in Fig. \ref{fig:stargw}. The waveform is also
similar and has the same property to the disk case.

%
\section{Discussion}
\label{sec:Discussion}
%
Using a BH perturbation approach, we discuss the size effect on GW
in the spectra from a dust disk and a spherical dust
star spiraling into a Kerr BH  for a wide range of parameters
($\tilde{L}_{z}$: spiraling case, $0 \leq a/M \leq 0.9$, $r_0=10M,
20M$).

First, we find that when the radius of a disk or a spherical star
is larger than the characteristic length of the QNM frequency, the
phase cancellation effect from the waves generated from each
different location of the particle appears in the spectrum.  Therefore,
it is meaningful to compare the diameter of the disk or the spherical
star with the characteristic length of GW, which is described by the
QNM frequency for the plunging orbit case.  We define the
characteristic length $L$ that the time delay between the earliest
particle at
$r_{0}-L$ and the latest one at $r_{0}+L$ is equivalent to the time
period of the characteristic frequency,
\begin{equation}
t(r_{0}+L) - t(r_{0}-L) = \frac{2 \pi}{\omega_{\rm QNM}}.
\label{eqn:length}
\end{equation}
Since the shape of the disk changes from the circle to a
long and narrow shape (Figs. \ref{fig:r10shapea9z2},
\ref{fig:r10shapea9z263}, and \ref{fig:r20shapea9z2}) and then
swallows into a BH, the definition of $L$ is indeed represents the
typical length scale of GW.  For the disk case, the time lag between
the earliest and the latest component particle plunging into a Kerr BH
is $\Delta t = t(r_{0}+R) - t(r_{0}-R)$.  We find that when $R$ is larger
than $L$, the energy spectrum of GW has a different feature from that
for a test particle.  This character comes from the phase cancellation
effect on GW from the earliest and the latest particle of the disk
plunging into a Kerr BH.  We summarize our result in Table
\ref{tbl:timelag}.  For the spherical star case, it is rather difficult to
argue with such a simple interpretation because the orbit has another
dependence $\theta$.  In spite of the complicated structure of the
geodesic for a spherical dust star, $C$ is almost constant in Tables
\ref{tbl:EstimateRs10} and \ref{tbl:EstimateRs20}, and we can
interpret them in a similar manner to the disk case.  We should also
argue the bound orbit case whether this phase cancellation effect
works in the inspiral phase of a coalescing binary system.  Although
the characteristic frequency is different from the QNM frequency, the
basic idea (phase cancellation effects on GW) would also be the same in
that case.  Therefore, the rule found in this paper can also apply to
the binary stars in the inspiral phase and it will be confirmed in the
future.

Next, we propose a possibility to determine the radius of a star.
The energy spectrum $(dE/d\omega)_{lm\omega}^{\rm (disk)}$ of GW
from a disk is expressed as
\begin{equation}
\left( \frac{dE}{d\omega} \right)_{lm\omega}^{\rm (disk)} \propto
\mid f_{m\omega}\mid ^2
\left( \frac{dE}{d\omega} \right)_{lm\omega}^{\rm (particle)},
\end{equation}
where $(dE/d\omega)_{lm\omega}^{\rm (particle)}$ is the spectrum 
from a single test particle.  The spectrum from a test particle has
only one peak at the frequency $\omega_{\rm QNM}$, so that the square
of the form factor $\mid f_{m\omega}\mid ^2$ is responsible for this
behavior.  The existence of several peaks in Figs. \ref{fig:r10a0z2f}
(c), \ref{fig:r10a5z2f} (c), \ref{fig:r10a9z2f} (c), \ref{fig:r10a9z263f}
(c), \ref{fig:r10a9z0f} (c), \ref{fig:r10a9zm3f} (c),  \ref{fig:r20a9z2f}
(b), (c) can be understood by the approximate estimation of
$f_{m\omega}$ as
\begin{equation}
 f_{m\omega} \propto \frac{\sin (\omega T'_{r=r_0} R)}{\omega},
\end{equation}
where $T'=dT(r)/dr$, assuming that only $e^{i \omega t}$ term
depends on $t$ in Eq. (\ref{eqn:FormFactor}).  The frequency where
$f_{m\omega}$ takes zero is
\begin{equation}
\omega_{n} \sim \frac{(n+1)\pi}{\ T'_{r=r_{0}} R},
\hspace{1cm}
n=0,1,2, \ldots ,
\label{eqn:EstimateF}
\end{equation}
which agrees quite well with the numerical results. Equation
(\ref{eqn:EstimateF}) suggests that the separation of peaks in the
spectrum $\Delta \omega$ may be in proportion to $R^{-1}$. In Table
\ref{tbl:EstimateR10} and \ref{tbl:EstimateR20}, we show
$\Delta \omega$ for various $R/M$ and we find the following relation
\begin{equation}
R=C\frac{1}{\Delta \omega}~ {\rm for}~ R \gg L.
\label{eqn:EstimateR}
\end{equation}
The value $C$ does not depend on the radius and take a similar value of
$\sim 1$, but depends on the orbits and the initial data, where we set
the location of the circular disk.  In fact, $C$ depends on them
within
$\sim 20\%$.  In the physical unit, $\Delta \nu\equiv \Delta
\omega/(2 \pi)$ is given by
\begin{equation}
\Delta \nu
= 5{\rm kHz} \left( \frac{R}{10{\rm km}} \right)^{-1}
= 8{\rm Hz} \left( \frac{R}{7000{\rm km}} \right)^{-1},
\end{equation}
assuming that $C=1$.  Therefore, for NSs and white dwarfs, the
frequency band is within the detectable frequency range of some laser
interferometers and resonant type detectors

Finally, we argue  from the observational point of view whether we
have a real situation to determine the radius of a tidally disrupted star
from GW in our model itself.  In the final phase of a coalescing binary
system, the effect of the tidal force and the deformation of the star
should be taken into account. Carter and Luminet \cite{CL83}
demonstrated the deformation of the star using an affine star model
with Newtonian particle dynamics (bound orbit).  They separate the
deformation stage into five phase, according to the position of the
star, and point out that when the star moves inside the Roche radius
with the condition of
$\beta
\equiv (R_{\rm Roche} / R_{\rm p} ) \gtrsim 3$, where $R_{\rm
Roche}$ is Roche radius and $R_{\rm p}$ is a periastron radius of the
orbit, the tidal force term rapidly grows and the pressure
and self gravity terms can be neglected.  This leads us to the
conclusion that we can neglect the effect of the pressure and the
self-gravity when $\beta \gtrsim 3$.  Laguna et al.
\cite{Laguna} extended the analysis of Carter and Luminet \cite{CL83}
using smooth particle hydrodynamics in Schwarzschild spacetime. 
They concluded that when $\beta \gtrsim 10$,  we can neglect the
effect of the self-gravity and the pressure.  From these previous
calculations, we may naturally set the model that a star is tidally
disrupted by a BH in a certain distance and then plunge into a BH.  In
this case, we may neglect the pressure force and the self-gravity
effect in the first approximation so that each of the component fluid
element follows the geodesic.

To apply our results to the realistic case,
three conditions should be satisfied.
  (1) The star is tidally disrupted at
$r_{\rm disrupt} \gtrsim 6M$,  which is expressed as
\begin{equation}
\frac{M}{r_{\rm disrupt}^{3}} \gtrsim
\rho_{\rm star} =
\frac{3 \mu}{4 \pi R^{3}},
\label{eqn:R1prev}
\end{equation}
where $\rho_{\rm star}$ is the density of the star.
(2) In order that the phase cancellation is effective
 in the spectrum, $R$ should be
larger than $L$.  This condition is described as
\begin{equation}
t(r_{0}+R) - t(r_{0}-R) > T = \frac{2 \pi}{\omega_{\rm QNM}}.
\label{eqn:R2prev}
\end{equation}
(3) The star does not contact to the BH.  This condition can be written as
\begin{equation}
R < r_{\rm disrupt} - r_{\rm horizon}.
\label{eqn:R3prev}
\end{equation}
We apply the NS of mass   $\mu = 1.4 M_{\odot}$ with  $R = 10 {\rm
km}$ and the white dwarf of mass $\mu = 0.5 M_{\odot}$ with $R =
7000 {\rm km}$ to our model. Then, two conditions [Eqs.
(\ref{eqn:R1prev}) and (\ref{eqn:R2prev})] are rewritten as
\begin{eqnarray}
M &\lesssim&
\sqrt{\frac{4 \pi R^{3}}{3 \mu}}
\left( \frac{r_{\rm disrupt}}{M} \right)^{-3/2}\equiv M_1
,\label{eqn:R1}
\\
M &\lesssim&
\frac{1}{\pi} \left. \frac{dt}{dr} \right|_{r=r_{0}}
R (M \omega_{\rm QNM})\equiv M_2
,
\label{eqn:R2}
\end{eqnarray}
where we use the first order approximation of Taylor expansion to
expand the left hand side of Eq. (\ref{eqn:R2prev}).  And the other 
condition [Eq. (\ref{eqn:R3prev})] gives the lower limit of the BH mass
as
\begin{eqnarray}
M &\gtrsim&
\frac{R}{
\left[
  \left( \frac{r_{\rm disrupt}}{M} \right) -
  \left( \frac{r_{\rm horizon}}{M} \right)
\right]
}\equiv M_3.
\label{eqn:R3}
\end{eqnarray}
Taking $r_{\rm disrupt}= 6M$
we tabulate these three conditions [Eqs. (\ref{eqn:R1}),
(\ref{eqn:R2}) and (\ref{eqn:R3})] in Tables
\ref{tbl:Region1} and \ref{tbl:Region2}.  From Tables \ref{tbl:Region1}
and \ref{tbl:Region2}, we indeed have a possibility to determine the
radius of the NS for   $M \sim 2
M_{\odot}$ and the radius of  the white dwarf for  $M \sim 1000 M_{\odot}$.

In the real situation, we should also take into account of the pressure
and self-gravity effect on GW so that it is urgent to confirm our
proposal by full 3D numerical simulations that we can indeed extract
the form factor from GW and determine the value of
$C$.  In any case, it is quite possible that the spectrum of GW may give
us important information in gravitational wave astronomy as in
optical astronomy.

\acknowledgments
M. S. thanks Masaru Shibata for discussion.   He also thanks the visitor
system of Yukawa Institute for Theoretical Physics and acknowledges
Stuart Shapiro at University of Illinois at Urbana-Champaign for his
hospitality.  The Numerical Computations are mainly performed by
NEC-SX vector computer at Yukawa Institute for Theoretical Physics,
Kyoto University and  FUJITSU-VX vector computer at Media Network
Center, Waseda University.  This work was supported in part by a JSPS
Grant-in-Aid (No. 095689) and by Grant-in-Aid of Scientific Research
of the Ministry of Education, Culture, and Sports, No.11640274 and
09NP0801.


%
\newpage
\figure
\begin{figure}
\caption{
Energy spectrum of gravitational waves from a disk moving on an
equatorial plane in Kerr spacetime whose radius is set up at $r=10M$
in the case of $a=0$, $\tilde{L}_{z} / M = 2$ [(a) $R=0$ (test particle),
(b) $R/M=0.785 $, (c) $R/M=2.33$, (d) $R/M=5.88$].  We only show $l=2$
mode. Solid, dashed, dash-dotted, dotted, dash-three dotted line
denotes the case of $m=-2$, $-1$, $0$, $1$, $2$, respectively.
}
\label{fig:r10a0z2sp}
\end{figure}

\begin{figure}
\caption{
Energy spectrum of gravitational waves from a disk moving on an
equatorial plane in Kerr spacetime whose radius is set up at $r=10M$
in the case of $a/M=0.5$, $\tilde{L}_{z} / M = 2$ [(a) $R=0$ (test
particle), (b) $R/M=0.785$, (c) $R/M=2.33$, (d) $R/M=5.88$].  We only
show $l=2$ mode. Solid, dashed, dash-dotted, dotted, dash-three
dotted line denotes the case of $m=-2$, $-1$, $0$, $1$, $2$,
respectively.}
\label{fig:r10a5z2sp}
\end{figure}

\begin{figure}
\caption{
Energy spectrum of gravitational waves from a disk moving on an
equatorial plane in Kerr spacetime whose radius is set up at $r=10M$
in the case of $a/M=0.9$, $\tilde{L}_{z} / M = 2$ [(a) $R=0$ (test
particle), (b) $R/M=0.785$, (c) $R/M=1.56$, (d) $R/M=5.88$].
We only show $l=2$ mode.  Solid, dashed, dash-dotted, dotted,
dash-three dotted line denotes the case of $m=-2$, $-1$, $0$, $1$, $2$,
respectively. }
\label{fig:r10a9z2sp}
\end{figure}

\begin{figure}
\caption{
Waveform of gravitational waves from a disk moving on an equatorial
plane in Kerr spacetime whose radius is set up at $r=10M$ in the case
of $a=0$, $\tilde{L}_{z} / M = 2$ [(a) $R=0$ (test particle), (b)
$R/M=0.785$, (c) $R/M=2.33$, (d) $R/M=5.88$].  We only show $l=m=2$
mode, setting the observer at the infinity of $\theta = \pi / 2$, $\phi =
0$. }
\label{fig:r10a0z2gw}
\end{figure}

\begin{figure}
\caption{
Waveform of gravitational waves from a disk moving on an equatorial
plane in Kerr spacetime whose radius is set up at $r=10M$ in the case
of $a/M=0.5$, $\tilde{L}_{z} / M = 2$ [(a) $R=0$ (test particle), (b)
$R/M=0.785$, (c) $R/M=2.33$, (d) $R/M=5.88$].  We only show $l=m=2$
mode, setting the observer at the infinity of $\theta = \pi / 2$, $\phi =
0$. }
\label{fig:r10a5z2gw}
\end{figure}

\begin{figure}
\caption{
Waveform of gravitational waves from a disk moving on an equatorial
plane in Kerr spacetime whose radius is set up at $r=10M$ in the case
of $a/M=0.9$, $\tilde{L}_{z} / M = 2$ [(a) $R=0$ (test particle), (b)
$R/M=0.785$, (c) $R/M=1.56$, (d) $R/M=5.88$].  We only show $l=m=2$
mode, setting the observer at the infinity of $\theta = \pi / 2$, $\phi =
0$. }
\label{fig:r10a9z2gw}
\end{figure}

\begin{figure}
\caption{
Form factor of the dust disk moving on the equatorial plane of a Kerr
spacetime whose radius is set up at $r=10M$ in the case of $a=0$,
$\tilde{L}_{z} / M = 2$ [(a) $R/M=0.785$, (b) $R/M=2.33$, (c)
$R/M=5.88$].  We only show the case for $l=m=2$.
}
\label{fig:r10a0z2f}
\end{figure}

\begin{figure}
\caption{
Form factor of the dust disk moving on the equatorial plane of a Kerr
spacetime whose radius is set up at $r=10M$ in the case of $a/M=0.5$,
$\tilde{L}_{z} / M = 2$ [(a) $R/M=0.785$, (b) $R/M=2.33$, (c)
$R/M=5.88$]. We only show the case for $l=m=2$.
}
\label{fig:r10a5z2f}
\end{figure}

\begin{figure}
\caption{
Form factor of the dust disk moving on the equatorial plane of a Kerr
spacetime whose radius is set up at $r=10M$ in the case of $a/M=0.9$,
$\tilde{L}_{z} / M = 2$ [(a) $R/M=0.785$, (b) $R/M=1.56$, (c)
$R/M=5.88$].  We only show the case for $l=m=2$.
}
\label{fig:r10a9z2f}
\end{figure}

\begin{figure}
\caption{
Energy spectrum of gravitational waves from a disk moving on an
equatorial plane in Kerr spacetime whose radius is set up at $r=10M$
in the case of $a/M=0.9$, $\tilde{L}_{z}/M = 2.63$ [(a) $R=0$ (test
particle), (b) $R/M=0.785$, (c) $R/M=1.56$, (d) $R/M=5.88$].  We only
show $l=2$ mode. Solid, dashed, dash-dotted, dotted, dash-three
dotted line denotes the case of $m=-2$, $-1$, $0$, $1$, $2$,
respectively. }
\label{fig:r10a9z263sp}
\end{figure}

\begin{figure}
\caption{
Energy spectrum of gravitational waves from a disk moving on an
equatorial plane in Kerr spacetime whose radius is set up at $r=10M$
in the case of $a/M=0.9$, $\tilde{L}_{z} = 0$ [(a) $R=0$ (test particle),
(b) $R/M=0.785$, (c) $R/M=1.56$, (d) $R/M=5.88$].  We only show $l=2$
mode.  Solid, dashed, dash-dotted, dotted, dash-three dotted line
denotes the case of $m=-2$, $-1$, $0$, $1$, $2$, respectively.
}
\label{fig:r10a9z0sp}
\end{figure}

\begin{figure}
\caption{
Energy spectrum of gravitational waves from a disk moving on an
equatorial plane in Kerr spacetime whose radius is set up at $r=10M$
in the case of $a/M=0.9$, $\tilde{L}_{z}/M = -3$ [(a) $R=0$ (test
particle), (b) $R/M=0.785$, (c) $R/M=1.56$, (d) $R/M=5.88$].  We only
show $l=2$ mode.  Solid, dashed, dash-dotted, dotted, dash-three
dotted line denotes the case of $m=-2$, $-1$, $0$, $1$, $2$,
respectively. }
\label{fig:r10a9zm3sp}
\end{figure}

\begin{figure}
\caption{
Waveform of gravitational waves from a disk moving on an equatorial
plane in Kerr spacetime whose radius is set up at $r=10M$ in the case
of $a/M=0.9$, $\tilde{L}_{z} / M = 2.63$ [(a) $R=0$ (test particle), (b)
$R/M=0.785$, (c) $R/M=1.56$, (d) $R/M=5.88$].  We only show $l=m=2$
mode, setting the observer at the infinity of $\theta = \pi / 2$, $\phi =
0$. }
\label{fig:r10a9z263gw}
\end{figure}

\begin{figure}
\caption{
Waveform of gravitational waves from a disk moving on an equatorial
plane in Kerr spacetime whose radius is set up at $r=10M$ in the case
of $a/M=0.9$, $\tilde{L}_{z} = 0$ [(a) $R=0$ (test particle), (b)
$R/M=0.785$, (c) $R/M=1.56$, (d) $R/M=5.88$].  We only show $l=m=2$
mode, setting the observer at the infinity of $\theta = \pi / 2$, $\phi =
0$. }
\label{fig:r10a9z0gw}
\end{figure}

\begin{figure}
\caption{
Waveform of gravitational waves from a disk moving on an equatorial
plane in Kerr spacetime whose radius is set up at $r=10M$ in the case
of $a/M=0.9$, $\tilde{L}_{z} / M = -3$ [(a) $R=0$ (test particle), (b)
$R/M=0.785$, (c) $R/M=1.56$, (d) $R/M=5.88$].  We only show $l=m=2$
mode, setting the observer at the infinity of $\theta = \pi / 2$, $\phi =
0$. }
\label{fig:r10a9zm3gw}
\end{figure}

\begin{figure}
\caption{
Form factor of the dust disk moving on the equatorial plane of a Kerr
spacetime whose radius is set up at $r=10M$ in the case of $a/M=0.9$,
$\tilde{L}_{z} / M = 2.63$ [(a) $R/M=0.785$, (b) $R/M=1.56$, (c)
$R/M=5.88$].  We only show the case for $l=m=2$.
}
\label{fig:r10a9z263f}
\end{figure}

\begin{figure}
\caption{
Form factor of the dust disk moving on the equatorial plane of a Kerr
spacetime whose radius is set up at $r=10M$ in the case of $a/M=0.9$,
$\tilde{L}_{z} = 0$ [(a) $R/M=0.785$, (b) $R/M=1.56$, (c) $R/M=5.88$].
We only show the case for $l=m=2$.
}
\label{fig:r10a9z0f}
\end{figure}

\begin{figure}
\caption{
Form factor of the dust disk moving on the equatorial plane of a Kerr
spacetime whose radius is set up at $r=10M$ in the case of $a/M=0.9$,
$\tilde{L}_{z} / M = -3$ [(a) $R/M=0.785$, (b) $R/M=1.56$, (c)
$R/M=5.88$]. We only show the case for $l=m=2$.
}
\label{fig:r10a9zm3f}
\end{figure}

\begin{figure}
\caption{
Deformation of the shape of the spherical disk whose radius is set up
at $r=10M$ in the case of $a/M=0.9$, $\tilde{L}_{z}/M=2$ [(a)
$R/M=1.56$, (b) $R/M=5.88$].  Solid line shows the geodesic for the
center of gravity of the disk, while circle, square, diamond, and
triangle show the edge of the disk where the location of the center is
at $r=10M$, $6M$, $4M$, $2M$, respectively.
}
\label{fig:r10shapea9z2}
\end{figure}

\begin{figure}
\caption{
Deformation of the shape of the spherical disk whose radius is set up
at $r=10M$ in the case of $a/M=0.9$, $\tilde{L}_{z}/M=2.63$ [(a)
$R/M=1.56$, (b) $R/M=5.88$].  Solid line shows the geodesic for the
center of gravity of the disk, while circle, square, diamond, and
triangle show the edge of the disk where the location of the center is
at $r=10M$, $6M$, $4M$, $2M$, respectively.
}
\label{fig:r10shapea9z263}
\end{figure}

\begin{figure}
\caption{
Deformation of the shape of the spherical disk whose radius is set up
at $r=20M$ in the case of $a/M=0.9$, $\tilde{L}_{z}/M=2$ [(a)
$R/M=3.13$, (b) $R/M=6.17$].  Solid line shows the geodesic for the
center of gravity of the disk, while circle, square, diamond, and
triangle show the edge of the disk where the location of the center is
at $r=20M$, $10M$, $4M$, $2M$, respectively.
}
\label{fig:r20shapea9z2}
\end{figure}

\begin{figure}
\caption{
Energy spectrum of gravitational waves from a disk moving on an
equatorial plane in Kerr spacetime whose radius is set up at $r=20M$
in the case of $a/M=0.9$, $\tilde{L}_{z} / M = 2$ [(a) $R/M=1.57$, (b)
$R/M=6.27$].  We only show $l=2$ mode.  Solid, dashed, dash-dotted,
dotted, dash-three dotted line denotes the case of $m=-2$, $-1$, $0$,
$1$, $2$, respectively. }
\label{fig:r20a9z2sp}
\end{figure}

\begin{figure}
\caption{
Waveform of gravitational waves from a disk moving on an equatorial
plane in Kerr spacetime whose radius is set up at $r=20M$ in the case
of $a/M=0.9$, $\tilde{L}_{z} / M = 2$ [(a) $R/M=1.57$, (b) $R/M=6.27$].
We only show $l=m=2$ mode, setting the observer at the infinity of
$\theta = \pi / 2$, $\phi = 0$.
}
\label{fig:r20a9z2gw}
\end{figure}

\begin{figure}
\caption{
Form factor of the dust disk moving on the equatorial plane of a Kerr
spacetime whose radius is set up at $r=20M$ in the case of $a/M=0.9$,
$\tilde{L}_{z} / M = 2$ [(a) $R/M=1.57$, (b) $R/M=6.27$].
We only show $l=m=2$ case.
}
\label{fig:r20a9z2f}
\end{figure}

\begin{figure}
\caption{
Spectrum of gravitational waves from a star in Kerr spacetime
in the case of $a/M=0.9$ [(a) $r_{0}=10M$, $R/M=1.56$, (b)
$r_{0}=10M$, $R/M=5.88$, (c) $r_{0}=20M$, $R/M=6.27$].  We only show
$l=2$ mode. Solid, dashed, dash-dotted, dotted, dash-three dotted line
denotes the case of $m=-2$, $-1$, $0$, $1$, $2$, respectively.
}
\label{fig:starsp}
\end{figure}

\begin{figure}
\caption{
Waveform of gravitational waves from a star in Kerr spacetime
in the case of $a/M=0.9$ [(a) $r_{0}=10M$, $R/M=1.56$, (b)
$r_{0}=10M$, $R/M=5.88$, (c) $r_{0}=20M$, $R/M=6.27$].  We only show
$l=m=2$ mode, setting the observer at the infinity of $\theta = \pi /
2$, $\phi = 0$. }
\label{fig:stargw}
\end{figure}

\newpage
\begin{table}
\caption
{
Comparison with the characteristic length which is appeared in energy
spectrum of gravitational waves to the radius of the disk in the case
of $r_{0}=10M$.  We only focus on $l=m=2$ mode. }
\begin{center}
\begin{tabular}{c c c c c c}
$a/M$ & $\tilde{L}/M$ & $R/M$ &  $M \Delta \omega$ &
$1/M \Delta \omega$  & $R \Delta \omega$\\
\hline
$0$ & $0$ &$3.09$ &  $0.370$ & $2.70$ & $1.14$
\\
$0$ & $0$ &$3.82$ &  $0.300$ & $3.33$ & $1.15$
\\
$0$ & $0$ &$4.54$ &  $0.255$ & $3.92$ & $1.16$
\\
$0$ & $0$ &$5.22$ &  $0.218$ & $4.59$ & $1.14$
\\
$0$ & $0$ &$5.88$ &  $0.193$ & $5.18$ & $1.13$
\\
$0$ & $2$ &$3.09$ &  $0.340$ & $2.94$ & $1.05$
\\
$0$ & $2$ &$3.83$ &  $0.275$ & $3.64$ & $1.05$
\\
$0$ & $2$ &$4.54$ &  $0.235$ & $4.26$ & $1.07$
\\
$0$ & $2$ &$5.22$ &  $0.203$ & $4.93$ & $1.06$
\\
$0$ & $2$ &$5.88$ &  $0.177$ & $5.65$ & $1.04$
\\
$0$ & $-3$ &$3.09$ &  $0.295$ & $3.39$ & $0.911$
\\
$0$ & $-3$ &$3.83$ &  $0.243$ & $4.12$ & $0.929$
\\
$0$ & $-3$ &$4.54$ &  $0.203$ & $4.93$ & $0.921$
\\
$0$ & $-3$ &$5.22$ &  $0.173$ & $5.78$ & $0.904$
\\
$0$ & $-3$ &$5.88$ &  $0.148$ & $6.76$ & $0.870$
\\
\hline
$0.5$ & $0$ &$3.09$ &  $0.340$ & $2.94$ & $1.05$
\\
$0.5$ & $0$ &$3.83$ &  $0.275$ & $3.64$ & $1.05$
\\
$0.5$ & $0$ &$4.54$ &  $0.235$ & $4.26$ & $1.07$
\\
$0.5$ & $0$ &$5.22$ & $0.203$ & $4.93$ & $1.06$
\\
$0.5$ & $0$ &$5.88$ &  $0.177$ & $5.65$ & $1.04$
\\
$0.5$ & $2$ &$3.09$ &  $0.335$ & $2.99$ & $1.03$
\\
$0.5$ & $2$ &$3.83$ &  $0.270$ & $3.70$ & $1.03$
\\
$0.5$ & $2$ &$4.54$ &  $0.233$ & $4.29$ & $1.06$
\\
$0.5$ & $2$ &$5.22$ &  $0.198$ & $5.05$ & $1.03$
\\
$0.5$ & $2$ &$5.88$ &  $0.177$ & $5.65$ & $1.04$
\\
$0.5$ & $-3$ &$3.09$ & $0.300$ & $3.33$ & $0.927$
\\
$0.5$ & $-3$ &$3.83$ & $0.248$ & $4.03$ & $0.949$
\\
$0.5$ & $-3$ &$4.54$ & $0.210$ & $4.76$ & $0.953$
\\
$0.5$ & $-3$ &$5.22$ & $0.177$ & $5.65$ & $0.924$
\\
$0.5$ & $-3$ &$5.88$ & $0.160$ & $6.25$ & $0.940$
\\
\hline
$0.9$ & $0$ &$3.09$ & $0.370$ & $2.70$ & $1.14$
\\
$0.9$ & $0$ &$3.83$ & $0.300$ & $3.33$ & $1.15$
\\
$0.9$ & $0$ &$4.54$ & $0.255$ & $3.92$ & $1.16$
\\
$0.9$ & $0$ &$5.22$ & $0.220$ & $4.55$ & $1.15$
\\
$0.9$ & $0$ &$5.88$ & $0.197$ & $5.08$ & $1.16$
\\
$0.9$ & $2$ &$3.09$ & $0.335$ & $2.99$ & $1.03$
\\
$0.9$ & $2$ &$3.83$ & $0.270$ & $3.70$ & $1.03$
\\
$0.9$ & $2$ &$4.54$ & $0.233$ & $4.29$ & $1.06$
\\
$0.9$ & $2$ &$5.22$ & $0.200$ & $5.00$ & $1.04$
\\
$0.9$ & $2$ &$5.88$ & $0.177$ & $5.65$ & $1.04$
\\
$0.9$ & $-3$ &$3.09$ & $0.310$ & $3.23$ & $0.957$
\\
$0.9$ & $-3$ &$3.83$ & $0.250$ & $4.00$ & $0.956$
\\
$0.9$ & $-3$ &$4.54$ & $0.213$ & $4.69$ & $0.967$
\\
$0.9$ & $-3$ &$5.22$ & $0.183$ & $5.46$ & $0.956$
\\
$0.9$ & $-3$ &$5.88$ & $0.163$ & $6.13$ & $0.958$
\\
\end{tabular}
\label{tbl:EstimateR10}
\end{center}
\end{table}

\begin{table}
\caption
{
Comparison with the characteristic length which is appeared in energy
spectrum of gravitational waves to the radius of the disk in the case
of $r_{0}=20M$.  We only focus on $l=m=2$ mode. }
\begin{center}
\begin{tabular}{c c c c c c}
$a/M$ & $\tilde{L}/M$ & $R/M$ &  $M \Delta \omega$ &
$1/M \Delta \omega$ & $R \Delta \omega$\\
\hline
$0$ & $0$ &$3.13$ &  $0.300$ & $3.33$ & $0.938$
\\
$0$ & $0$ &$4.67$ &  $0.192$ & $5.21$ & $0.896$
\\
$0$ & $0$ &$6.18$ &  $0.154$ & $6.49$  & $0.951$
\\
$0$ & $2$ &$3.13$ &  $0.290$ & $3.45$ & $0.907$
\\
$0$ & $2$ &$4.67$ &  $0.182$ & $5.49$ & $0.849$
\\
$0$ & $2$ &$6.18$ & $0.146$ & $6.85$ & $0.902$
\\
$0$ & $-3$ &$3.13$ & $0.270$ & $3.70$ & $0.844$
\\
$0$ & $-3$ &$4.67$ & $0.172$ & $5.81$ & $0.803$
\\
$0$ & $-3$ &$6.18$ &  $0.130$ & $7.69$  & $0.803$
\\
\hline
$0.5$ & $0$ &$3.13$ &  $0.300$ & $3.33$ & $0.938$
\\
$0.5$ & $0$ &$4.67$ &  $0.188$ & $5.32$ & $0.877$
\\
$0.5$ & $0$ &$6.18$ & $0.141$ & $7.09$ & $0.871$
\\
$0.5$ & $2$ &$3.14$ & $0.290$ & $3.45$ & $0.907$
\\
$0.5$ & $2$ &$4.67$ &  $0.183$ & $5.46$ & $0.854$
\\
$0.5$ & $2$ &$6.18$ &  $0.138$ & $7.25$ & $0.852$
\\
$0.5$ & $-3$ &$3.14$ & $0.275$ & $3.64$ & $0.860$
\\
$0.5$ & $-3$ &$4.71$ & $0.172$ & $5.81$ & $0.803$
\\
$0.5$ & $-3$ &$6.27$ & $0.127$ & $7.87$ & $0.785$
\\
\hline
$0.9$ & $0$ &$3.13$ & $0.300$ & $3.33$ & $0.938$
\\
$0.9$ & $0$ &$4.67$ & $0.195$ & $5.13$ & $0.910$
\\
$0.9$ & $0$ &$6.17$ & $0.141$ & $7.09$ & $0.871$
\\
$0.9$ & $2$ &$3.13$ & $0.285$ & $3.51$ & $0.891$
\\
$0.9$ & $2$ &$4.67$ & $0.185$ & $5.41$ & $0.863$
\\
$0.9$ & $2$ &$6.17$ & $0.133$ & $7.52$ & $0.822$
\\
$0.9$ & $-3$ &$3.13$ & $0.275$ & $3.64$ & $0.860$
\\
$0.9$ & $-3$ &$4.67$ & $0.173$ & $5.78$ & $0.807$
\\
$0.9$ & $-3$ &$6.17$ & $0.150$ & $6.67$ & $0.927$
\\
\end{tabular}
\label{tbl:EstimateR20}
\end{center}
\end{table}

\begin{table}
\caption
{
Comparison with the characteristic length which is appeared in energy
spectrum of gravitational waves to the radius of the star in the case
of $r_{0}=10M$.  We only focus on $l=m=2$ mode. }
\begin{center}
\begin{tabular}{c c c c c}
$a/M$ & $R/M$ &  $M \Delta \omega$ &
$1/M \Delta \omega$ & $R \Delta \omega$\\
\hline
$0$ &$5.22$ &  $0.175$ & $5.71$ & $0.914$
\\
$0$ &$5.88$ &  $0.145$ & $6.90$ & $0.852$
\\
$0.5$ &$5.22$ &  $0.175$ & $5.71$ & $0.914$
\\
$0.5$ &$5.88$ & $0.145$ & $6.90$ & $0.852$
\\
$0.9$ &$5.22$ &  $0.170$ & $5.88$ & $0.888$
\\
$0.9$ &$5.88$ & $0.150$ & $6.67$ & $0.881$
\\
\end{tabular}
\label{tbl:EstimateRs10}
\end{center}
\end{table}

\begin{table}
\caption
{
Comparison with the characteristic length which is appeared in energy
spectrum of gravitational waves to the radius of the star in the case
of $r_{0}=20M$.  We only focus on $l=m=2$ mode. }
\begin{center}
\begin{tabular}{c c c c c}
$a/M$ & $R/M$ &  $M \Delta \omega$ &
$1/M \Delta \omega$ & $R \Delta \omega$\\
\hline
$0$ &$6.18$ &  $0.260$ & $3.85$ & $1.61$
\\
$0.5$ &$4.67$ &  $0.285$ & $3.51$ & $1.33$
\\
$0.5$ &$6.18$ & $0.250$ & $4.00$ & $1.54$
\\
$0.9$ &$4.67$ &  $0.260$ & $3.85$ & $1.21$
\\
$0.9$ &$6.18$ & $0.245$ & $4.08$ & $1.51$
\\
\end{tabular}
\label{tbl:EstimateRs20}
\end{center}
\end{table}

\begin{table}
\caption
{
Characteristic length of gravitational waves for a plunging orbit.  We
define the characteristic length $L$ with satisfying Eq.
(\ref{eqn:length}).  We solve geodesic equation numerically in order to
find the characteristic length.  We only use $l=m=2$ mode of the QNM
frequency.}
\begin{center}
\begin{tabular}{c c c c}
$a/M$ & $\tilde{L}_{z}/M$ &
$L_{r_{0}=10M}$ & $L_{r_{0}=20M}$\\
\hline
$0$ &$0$ &  $2.78$ &  $2.31$
\\
$0$ &$2$ &  $3.03$ &  $2.42$
\\
$0$ &$-3$ & $2.42$ &  $2.16$
\\
$0.5$ &$0$ &  $2.22$ &  $1.85$
\\
$0.5$ &$2$ & $2.44$ &  $1.94$
\\
$0.5$ &$-3$ & $1.99$ &  $1.74$
\\
$0.9$ &$0$ &  $1.54$ &  $1.29$
\\
$0.9$ &$2$ & $1.71$ &  $1.36$
\\
$0.9$ &$-3$ & $1.42$ &  $1.22$
\\
\end{tabular}
\label{tbl:timelag}
\end{center}
\end{table}

\begin{table}
\caption
{
Allowed mass of the BH (Eqs. (\ref{eqn:R1}) and (\ref{eqn:R2}))
for a NS ($\mu = 1.4 M_{\odot}$, $R=10 {\rm km}$) to determine the
radius of the star.  $M_{1}$ and $M_{2}$ denote the upper limit of the
mass from Eqs. (\ref{eqn:R1}) and (\ref{eqn:R2}) and $M_{3}$ denote
the lower limit of the mass from Eq. (\ref{eqn:R3}), respectively.}
\begin{center}
\begin{tabular}{c c c c c}
$a/M$ & $\tilde{L}_{z}/M$ &
$M_{1}/M_{\odot}$ & $M_{2}/M_{\odot}$ & $M_{3}/M_{\odot}$\\
\hline
$0$ &$0$ &  $2.07$ & $2.07$ & $1.69$
\\
$0$ &$2$ &  $2.07$ & $2.06$ & $1.69$
\\
$0$ &$-3$ & $2.07$ & $2.05$ & $1.69$
\\
$0.5$ &$0$ &  $2.07$ & $2.55$ & $1.64$
\\
$0.5$ &$2$ & $2.07$ & $2.75$ & $1.64$
\\
$0.5$ &$-3$ & $2.07$ & $2.28$ & $1.64$
\\
$0.9$ &$0$ &  $2.07$ & $3.56$ & $1.48$
\\
$0.9$ &$2$ & $2.07$ & $4.12$ & $1.48$
\\
$0.9$ &$-3$ & $2.07$ & $3.00$ & $1.48$
\\
\end{tabular}
\label{tbl:Region1}
\end{center}
\end{table}

\begin{table}
\caption
{
Allowed mass of BH (Eqs. (\ref{eqn:R1}) and (\ref{eqn:R2}))
for a white dwarf ($\mu = 0.5 M_{\odot}$, $R=7,000 {\rm km}$) to
determine the radius of the star. 
$M_{1}$ and $M_{2}$ denote the upper limit of the mass from Eqs.
(\ref{eqn:R1}) and (\ref{eqn:R2}) and $M_{3}$ denote the lower limit
of the mass from Eq. (\ref{eqn:R3}), respectively.}
\begin{center}
\begin{tabular}{c c c c c}
$a/M$ & $\tilde{L}_{z}/M$ &
$M_{1}/M_{\odot}$ & $M_{2}/M_{\odot}$ & $M_{3}/M_{\odot}$\\
\hline
$0$ &$0$ &  $64300$ & $1450$ & $1180$
\\
$0$ &$2$ &  $64300$ & $1440$ & $1180$
\\
$0$ &$-3$ & $64300$ & $1440$ & $1180$
\\
$0.5$ &$0$ &  $64300$ & $1780$ & $1150$
\\
$0.5$ &$2$ & $64300$ & $1920$ & $1150$
\\
$0.5$ &$-3$ & $64300$ & $1600$ & $1150$
\\
$0.9$ &$0$ &  $64300$ & $2490$ & $1040$
\\
$0.9$ &$2$ & $64300$ & $2880$ & $1040$
\\
$0.9$ &$-3$ & $64300$ & $2100$ & $1040$
\\
\end{tabular}
\label{tbl:Region2}
\end{center}
\end{table}


\begin{thebibliography}{99}
\bibitem{ODE}
If we only want to consider the merging phase of a coalescing
binary BH - star system, it is very useful to evolve the Teukolsky
equation with setting a suitable initial data (Close limit approximation
\cite{CL}). In this case, we have to compute a set of partially
differential equations.  This may guide the result with less
accuracy.  We also know the result in the case of a test particle that
most of gravitational waves are radiated in the merging phase.  In our
approach, we only need the analysis of Fourier transformation
method with calculating the set of ordinary differential equations.
\end{thebibliography}
\end{document}